\documentclass[12pt]{article}%
\usepackage{amsmath}
\usepackage{amsfonts}
\usepackage{amssymb}
\usepackage{graphicx}%
\providecommand{\U}[1]{\protect \rule{.1in}{.1in}}
\newtheorem{theorem}{Theorem}

\newtheorem{proposition}[theorem]{Proposition}
\newtheorem{remark}[theorem]{Remark}

\newenvironment{proof}[1][Proof]{\noindent \textbf{#1.} }{\  \rule{0.5em}{0.5em}}
\begin{document}

\title{Expanding universes in the conformal frame of $f\left(  R\right)  $ gravity\footnote{%
Talk given in \textquotedblleft The Invisible Universe\textquotedblright ,
June 29 - July 3, 2009, Paris.}}

\author{John Miritzis,\\Department of Marine Sciences, University of the Aegean,\\University Hill, Mytilene 81100, Greece\\and\\Roberto Giamb\`{o}\\Department of Mathematics and Computer Science \\University of Camerino 62032 Camerino (MC) Italy}
\maketitle

\begin{abstract}
The late time evolution of Friedmann-Robertson-Walker (FRW) models with a
perfect fluid matter source is studied in the conformal frame of $f\left(
R\right) $ gravity. We assume that the corresponding scalar field,
nonminimally coupled to matter, has an arbitrary non-negative potential
function $V\left( \phi \right) $. We prove that equilibria corresponding to
non-negative local minima for $V$ are asymptotically stable. We investigate
all cases where one of the matter components eventually dominates. The
results are valid for a large class of non-negative potentials without any
particular assumptions about the behavior of the potential at infinity. In
particular for a nondegenerate minimum of the potential with zero critical
value we show that if $\gamma $, the parameter of the equation of state is
larger than one, then there is a transfer of energy from the fluid to the
scalar field and the later eventually dominates.
\end{abstract}
\maketitle
\section{Introduction}

The standard inflationary idea requires that there be a period of slow-roll
evolution of a scalar field (the inflaton) during which its potential energy
drives the universe in a quasi-exponential expansion. Besides a cosmological
constant, a massive scalar field (quintessence) provides the simplest
mechanism to obtain accelerated expansion of the universe within General
Relativity. Therefore, scalar fields play a prominent role in the
construction of cosmological scenarios aiming to describe the evolution of
the early and the present universe. Since the nature of the scalar field
supposed to cause accelerated expansion is unknown, it is important to
investigate the general properties shared by all FRW models with a scalar
field irrespectively of the particular choice of the potential.

In this paper we study the late time evolution of initially expanding flat
and negatively curved FRW models with a scalar field having an arbitrary
bounded from below potential function $V\left( \phi \right) $. The scalar
field is nonminimally coupled to ordinary matter described by a barotropic
fluid with equation of state 
\begin{equation*}
p=(\gamma -1)\rho ,\ \ \ 0<\gamma \leq 2.
\end{equation*}%
The corresponding case of initially contracting models was studied in \cite%
{giam2}. Nonminimally coupling occurs for example, in string theory \cite%
{gasp}, in scalar-tensor theories of gravity \cite{fuma}, in higher order
gravity (HOG) theories \cite{cafa} and in models of chameleon gravity \cite%
{wate}. In particular, for HOG theories derived from Lagrangians of the form 
\begin{equation}
L=f\left( R\right) \sqrt{-g}+2L_{\mathrm{m}}\left( \Psi \right) ,
\label{lagr}
\end{equation}%
it is well known that under a suitable conformal transformation the field
equations reduce to the Einstein field equations with a scalar field $\phi $
as an additional matter source.

The conformal equivalence can be formally obtained by conformally
transforming the Lagrangian (\ref{lagr}) and the resulting action becomes 
\cite{bbpst}, 
\begin{equation*}
\widetilde{S}=\int d^{4}x\sqrt{-\widetilde{g}}\left\{ \widetilde{R}-\left[
\left( \partial\phi\right) ^{2}+2V\left( \phi\right) \right] +2e^{-2\sqrt{2/3%
}\phi}\mathcal{L}_{\mathrm{m}}\left( e^{-\sqrt{2/3}\phi }\widetilde{g}%
,\Psi\right) \right\} .
\end{equation*}
Variation of $\widetilde{S}$ with respect to $\widetilde{g}$ yields the
field equations, 
\begin{equation}
\widetilde{G}_{\mu\nu}=T_{\mu\nu}\left( \widetilde{g},\phi\right) +%
\widetilde{T}_{\mu\nu}\left( \widetilde{g},\Psi\right) ,  \label{confm}
\end{equation}
and variation of $\widetilde{S}$ with respect to $\phi$ yields the equation
of motion of the scalar field,%
\begin{equation}
\widetilde{\square}\phi-\frac{dV}{d\phi}=\frac{1}{\sqrt{6}}e^{-2\sqrt{2/3}%
\phi}T_{\mu}^{\mu}\left( \widetilde{g},\Psi\right) .  \label{emsf}
\end{equation}
Note that the Bianchi identities imply that 
\begin{equation*}
\widetilde{\nabla}^{\mu}\widetilde{T}_{\mu\nu}\left( \widetilde{g}%
,\Psi\right) \neq0,\ \ \ \ \widetilde{\nabla}^{\mu}T_{\mu\nu}\left( 
\widetilde{g},\phi\right) \neq0, 
\end{equation*}
and therefore there is an energy exchange between the scalar field and
ordinary matter.

\section{Flat and negatively curved FRW with an arbitrary non-negative
potential}

For homogeneous and isotropic spacetimes the field equations (\ref{confm})
reduce to the Friedmann equation, 
\begin{equation}
H^{2}+\frac{k}{a^{2}}=\frac{1}{3}\left( \rho +\frac{1}{2}\dot{\phi}%
^{2}+V\left( \phi \right) \right) ,  \label{fri1jm}
\end{equation}%
and the Raychaudhuri equation, 
\begin{equation}
\dot{H}=-\frac{1}{2}\dot{\phi}^{2}-\frac{\gamma }{2}\rho +\frac{k}{a^{2}},
\label{fri2jm}
\end{equation}%
while the equation of motion of the scalar field (\ref{emsf}), becomes 
\begin{equation}
\ddot{\phi}+3H\dot{\phi}+V^{\prime }\left( \phi \right) =\frac{4-3\gamma }{%
\sqrt{6}}\rho .  \label{emsjm}
\end{equation}%
The Bianchi identities yield the conservation equation, 
\begin{equation}
\dot{\rho}+3\gamma \rho H=-\frac{4-3\gamma }{\sqrt{6}}\rho \dot{\phi},
\label{conssfjm}
\end{equation}%
(see for example \cite{bbpst,maso}). For simplicity we drop the tilde from
all quantities. We adopt the metric and curvature conventions of \cite{wael}%
. $a\left( t\right) $ is the scale factor, an overdot denotes
differentiation with respect to time $t,$ $H=\dot{a}/a$ and units have been
chosen so that $c=1=8\pi G$. Here $V\left( \phi \right) $ is the potential
energy of the scalar field and $V^{\prime }=dV/d\phi $.

In the following we consider initially expanding solutions of (\ref{fri2jm}%
)-(\ref{conssfjm}), i.e. $H(0)>0$. For flat, $k=0,$ models the state vector
of the system (\ref{fri2jm})-(\ref{conssfjm}) is $\left( \phi ,\dot{\phi}%
,\rho ,H\right) $, i.e. we have a four-dimensional dynamical system subject
to the constraint (\ref{fri1jm}). Defining $y:=\dot{\phi}\ $and setting $%
(4-3\gamma )/\sqrt{6}=:\alpha ,$ we write the autonomous system as%
\begin{align}
\dot{\phi}& =y,  \notag \\
\dot{y}& =-3Hy-V^{\prime }\left( \phi \right) +\alpha \rho ,  \notag \\
\dot{\rho}& =-3\gamma \rho H-\alpha \rho y,  \label{sys1} \\
\dot{H}& =-\frac{1}{2}y^{2}-\frac{\gamma }{2}\rho ,  \notag
\end{align}%
subject to the constraint 
\begin{equation}
3H^{2}=\rho +\frac{1}{2}y^{2}+V\left( \phi \right) .  \label{cons1}
\end{equation}

\begin{remark}
The function $W$\ defined by 
\begin{equation*}
W(\phi ,y,\rho ,H)=H^{2}-\frac{1}{3}\left( \frac{1}{2}y^{2}+V(\phi )+\rho
\right) ,  
\end{equation*}%
satisfies 
\begin{equation*}
\dot{W}=-2HW,  
\end{equation*}%
and therefore, the hypersurface $\{W=0\}$, representing flat, $k=0$,
cosmologies is invariant under the flow of \eqref{sys1}. By standard
arguments in ordinary differential equations theory it follows that also $%
sgn(W)$\ is invariant under the flow of \eqref{sys1}. We deduce that
solutions with $W$\ positive, null, or negative, represent scalar field
cosmologies with $k=-1,0,1$\ respectively. Similar arguments applied to the
third of \eqref{sys1} show that if $\rho >0$\ at some initial time $t_{0}$,
then $\rho (t)>0$\ throughout the solution. Denoting by 
\begin{equation*}
\epsilon =\frac{1}{2}y^{2}+V\left( \phi \right) ,
\end{equation*}%
the energy density of the scalar field, we see that 
\begin{equation*}
\dot{\epsilon}+\dot{\rho}=-3H(y^{2}+\gamma \rho ),  
\end{equation*}%
which implies that, for expanding models the total energy $\epsilon +\rho $\
of the system decreases.\end{remark}

The equilibria of \eqref{sys1} are given by $(\phi=\phi_{\ast},y=0,\rho
=0,H=\pm\sqrt{V(\phi_{\ast})/3})$ where $V^{\prime}(\phi_{\ast})=0$.
Regarding their stability for expanding cosmologies, we note the following
facts. Critical points of $V$ with negative critical value are not
equilibria and they rather allow for recollapse of the model. Moreover,
nondegenerate maximum points (with non negative critical value) for $V$ are
unstable, as can be easily seen by linearizing system \eqref{sys1} at the
corresponding equilibria and verifying the existence of at least one
eigenvalue with positive real part. It is interesting to study what happens
near local minima of the potential with non negative critical value, and a
stability result can be given when $k=0,-1$.

\begin{proposition}
Let $\phi _{\ast }$ a strict local minimum for $V(\phi )$, possibly
degenerate, with nonnegative critical value. Then, $\mathbf{p}_{\ast }=(\phi
_{\ast },y_{\ast }=0,\rho _{\ast }=0,H_{\ast }=\sqrt{\tfrac{V(\phi _{\ast })%
}{3}})$ is an asymptotically stable equilibrium point for expanding
cosmologies in the open spatial topologies $k=0$ and $k=-1$.
\end{proposition}

\begin{proof}[Sketch of the proof]
The proof consists in constructing a compact set $\Omega$ in $\mathbb{R}^{4}$
and showing that it is positively invariant. Applying LaSalle's invariance
theorem to the functions $W$ and $(\rho+\epsilon)$ in $\Omega$, it is shown
that every trajectory in $\Omega\ $is such that $HW\rightarrow0$ and $%
H(y^{2}+\gamma\rho)\rightarrow0$ as $t\rightarrow+\infty$, which means $%
y\rightarrow0,\,\rho\rightarrow0$, and $H^{2}-\tfrac{1}{3}V(\phi
)\rightarrow0$. Since $H$ is monotone and admits a limit, $V(\phi)$ also
admits a limit, $V(\phi_{\ast})$, thus the solution approaches the
equilibrium point $\mathbf{p}_{\ast}$. Details in \cite{gimi}.
\end{proof}

Similar results were proved in \cite{miri2} for separately conserved scalar
field and perfect fluid.

\section{Energy exchange}

In the following we restrict ourselves to the flat, $k=0,$ case and study
the energy transfer from the perfect fluid to the scalar field. We are
interested to study the late time behavior near the equilibrium point $(\phi
=\phi _{\ast },y=0,\rho =0,H=\sqrt{\tfrac{V(\phi _{\ast })}{3}})$, which, by
the previous Proposition is asymptotically stable. We suppose that the
initial data in the basin of attraction of this equilibrium are such that
the fluid is the dominant matter component, i.e. 
\begin{equation*}
\rho _{0}>\epsilon _{0},
\end{equation*}%
and we investigate whether there is a time $t_{1}$ such that 
\begin{equation}
\epsilon \left( t\right) >\rho (t),\qquad \forall t>t_{1}.
\label{eq:transition}
\end{equation}%
This question is relevant to the coincidence problem, that is, why dark
energy and matter appear to have roughly the same energy density today (see 
\cite{aqtw} and references therein).

If $V\left( \phi _{\ast }\right) >0,$ the transition (\ref{eq:transition})
does always happen. In fact, it easily follows that if the critical value of
the potential is strictly positive, it behaves as an effective cosmological
constant and the energy of the scalar field tends to this value whereas the
energy of the fluid tends to zero. We conclude that the relevant case is
when $V(\phi _{\ast })=0$ and hereafter we will focus on the case when $\phi
_{\ast }$ is a nondegenerate minimum. Without loss of generality we suppose
that $\phi _{\ast }=0$ and therefore, the general form of the potential
studied can be written in a neighborhood of $\phi =0$ as%
\begin{equation}
V(\phi )=\frac{1}{2}\omega ^{2}\phi ^{2}+\mathcal{O}(\phi ^{3}),\qquad
\omega >0.  \label{eq:V}
\end{equation}
Integrating (\ref{conssfjm}), we get 
\begin{equation}
\rho(t)=ce^{-\alpha\phi(t)}a(t)^{-3\gamma}.  \label{eq:rho-as}
\end{equation}
From the Proposition in the previous Section we know \emph{a-priori} that $%
\left( \phi,\dot{\phi},\rho,H\right) \rightarrow\left( 0,0,0,0\right) $ as $%
t\rightarrow\infty$, hence we can write $\rho(t)\simeq ca(t)^{-3\gamma}$ as $%
t\rightarrow\infty$.

We now turn our attention to the equation of motion of the scalar field (\ref%
{emsjm}) with the potential \eqref{eq:V}, namely 
\begin{equation}
\ddot{\phi}+3H\dot{\phi}+\omega ^{2}\phi +\mathcal{O}(\phi ^{2})=\alpha \rho
.  \label{emsf1}
\end{equation}%
This equation can be solved by the Kryloff-Bogoliuboff (KB) approximation 
\cite{krbo}. We present an outline of the method for the convenience of
readers with no previous knowledge of the KB approximation. Consider the
differential equation 
\begin{equation}
\ddot{\phi}+\eta f\left( \phi ,\dot{\phi}\right) +\omega ^{2}\phi =0,\ \ \ \
0<\eta \ll 1.  \label{de}
\end{equation}%
If $\eta =0,$ the solution can be written as 
\begin{equation*}
\phi \left( t\right) =\frac{1}{\omega }r\cos \left( -\omega t+\chi \right) \
\ \ \ \mathrm{and\ }\ \ \dot{\phi}\left( t\right) =r\sin \left( -\omega
t+\chi \right) ,
\end{equation*}%
where $r$ and $\chi $ are arbitrary constants. We are looking for a solution
of (\ref{de}) which resembles to the form of the simple harmonic oscillator,
that is, 
\begin{equation}
\phi \left( t\right) =\frac{1}{\omega }r\left( t\right) \cos \left( -\omega
t+\chi \left( t\right) \right) \ \ \ \ \mathrm{and\ }\ \ \dot{\phi}\left(
t\right) =r\left( t\right) \sin \left( -\omega t+\chi \left( t\right)
\right) .  \label{harm}
\end{equation}%
Setting $\theta \left( t\right) =-\omega t+\chi \left( t\right) $ and
substituting (\ref{harm}) in (\ref{de}) yields 
\begin{align}
\frac{dr}{dt}& =-{\eta }f\left( \frac{1}{\omega }r\cos \theta ,r\sin \theta
\right) \sin \theta ,  \label{afi} \\
\frac{d\chi }{dt}& =-\frac{\eta }{r}f\left( \frac{1}{\omega }r\cos \theta
,r\sin \theta \right) \cos \theta .  \notag
\end{align}%
For $\eta $ small, $r(t)$ and $\chi (t)$ are slowly varying functions of $t.$
Consequently, in a time $T=2\pi /\omega ,$ $r(t)$ and $\chi (t)$ have not
changed appreciably, while $\theta \left( t\right) =-\omega t+\chi \left(
t\right) $ will increase approximately by $-2\pi .$ Therefore, we replace
the rhs of (\ref{afi}) by their average values over a range of $2\pi $ of $%
\theta $, i.e. the amplitude $r(t)$ is\emph{\ regarded as a constant} in
taking the average (this is the essence of the KB approximation). This leads to the equations
\begin{align*}
\frac{dr}{dt}& =-\frac{\eta }{2\pi }\int_{0}^{2\pi }f\left( \frac{1}{\omega }%
r\cos \theta ,r\sin \theta \right) \sin \theta d\theta , \\
\frac{d\chi }{dt}& =-\frac{\eta }{2\pi r}\int_{0}^{2\pi }f\left( \frac{1}{%
\omega }r\cos \theta ,r\sin \theta \right) \cos \theta d\theta .
\end{align*}
If we could apply the KB approximation to (\ref{emsf1}) with 
\begin{equation}
\eta f\left( \phi ,\dot{\phi}\right) =3H\dot{\phi}+\mathcal{O}(\phi
^{2})-\alpha \rho ,  \label{eq:eta}
\end{equation}%
then we should obtain for the amplitude of $\phi $%
\begin{equation}
\frac{dr}{dt}=-\frac{1}{2\pi }\int_{0}^{2\pi }\left( 3H\omega r\cos \theta +%
\mathcal{O}(r^{2}\cos ^{2}\theta )-\alpha ce^{-(\alpha /\omega )r\cos \theta
}a^{-3\gamma }\right) \sin \theta d\theta .  \label{ampl}
\end{equation}%
The integral 
\begin{equation*}
\int_{0}^{2\pi }e^{-(\alpha /\omega )r\cos \theta }\sin \theta d\theta ,
\end{equation*}%
vanishes and since $H$ is a slow varying function compared to the rapid
oscillations of the scalar field, it can be considered as constant during
one period, $T=2\pi /\omega $. Therefore, (\ref{ampl}) becomes%
\begin{equation}
\frac{dr}{dt}=-\frac{3}{2}Hr+\mathcal{O}(r^{3}),  \label{ampl1}
\end{equation}%
since the integration of the quadratic terms vanish. Ignoring third-order
terms, eq. (\ref{ampl1}) can be integrated to give $r(t)=r_{0}(t)a^{-3/2}$,
where $r_{0}(t)$ is bounded and bounded away from zero uniformly as $%
t\rightarrow +\infty $.

Hence we find that for large $t$ the amplitude of the scalar field varies as 
$a^{-3/2}$. Since the amplitude of $\dot{\phi}$ has by (\ref{harm}) the same
time dependence as the amplitude of $\phi$ and in our approximation 
\begin{equation*}
\epsilon\cong\frac{1}{2}\dot{\phi}^{2}+\frac{1}{2}\omega^{2}\phi^{2},
\end{equation*}
it follows that 
\begin{equation*}
\epsilon\cong a^{-3}.
\end{equation*}
Comparing this result with the time dependence of the fluid density $%
\rho\cong a^{-3\gamma}$, we arrive at the following picture for the
evolution of the universe. \medskip

\emph{If }$\gamma<1$\emph{\ the energy density }$\rho$\emph{\ eventually
dominates over the energy density of the scalar field }$\epsilon$\emph{\ and
this universe follows the classical Friedmannian evolution. For }$\gamma >1$%
\emph{, since }$\rho$\emph{\ decreases faster than the energy density }$%
\epsilon$\emph{\ of the scalar field, }$\epsilon$\emph{\ eventually
dominates over }$\rho$.\medskip

A rigorous proof of these results is given in \cite{gimi}. The main
obstruction to apply the KB approximation is due to \eqref{eq:eta}, indeed,
the above argument applies if $\rho/r$ goes to zero. Otherwise, the rhs in %
\eqref{afi} is not infinitesimal and so $\chi(t)$ could in principle be
comparable to $\omega t$.

Our results could be interesting in investigations of cosmological scenarios
in which the energy density of the scalar field mimics the background energy
density. For viable dark energy models, it is necessary that the energy
density of the scalar field remains insignificant during most of the history
of the universe and emerges only at late times to account for the current
acceleration of the universe.

\end{document}